\documentclass[aps,prd,twocolumn,superscriptaddress]{revtex4}
\usepackage{epsfig,epsf}
\usepackage{amsmath}
\usepackage{amsthm}
\usepackage{amsfonts}
\usepackage{amssymb}
\usepackage{dsfont}
\usepackage{multirow}
\usepackage{appendix}
\usepackage{slashed}
\usepackage[active]{srcltx}
\usepackage{psfrag}

\begin{document}

\title{Tensor hybrid mesons $\overline{b}gc$ }
\date{\today}
\author{S.~S.~Agaev}
\affiliation{Institute for Physical Problems, Baku State University, Az--1148 Baku,
Azerbaijan}
\author{K.~Azizi}
\thanks{Corresponding Author}
\affiliation{Department of Physics, University of Tehran, North Karegar Avenue, Tehran
14395-547, Iran}
\affiliation{Department of Physics, Dogus University, Dudullu-\"{U}mraniye, 34775
Istanbul, T\"{u}rkiye}
\author{H.~Sundu}
\affiliation{Department of Physics Engineering, Istanbul Medeniyet University, 34700
Istanbul, T\"{u}rkiye}

\begin{abstract}
Spectroscopic parameters and widths of the tensor hybrid mesons $H_{\mathrm{%
bc}}$ and $\widetilde{H}_{\mathrm{bc}}$ with the structure $\overline{b}gc$
and spin-parities $J^{\mathrm{P}}=2^{-}$ and $J^{\mathrm{P}}=2^{+}$ are
calculated with high accuracy in the QCD sum rule framework. Information on
their masses $m=(7.214\pm 0.075)~\mathrm{GeV}$ and $\widetilde{m} =(7.685\pm
0.040)~\mathrm{GeV}$ enable us to determine decay channels of $H_{\mathrm{bc}%
}$ and $\widetilde{H}_{\mathrm{bc}}$. The full width of the meson $H_{%
\mathrm{bc}}$ is estimated by considering the decays $H_{\mathrm{bc}} \to
D^{+}\overline{B}^{\ast 0}$ and $D^{0}B^{\ast +}$. The channels $\widetilde{H%
}_{\mathrm{bc}}\to B^{+}D^{0}$, $B^{0}D^{+}$, $B^{\ast +}D^{\ast 0}$, $%
B^{\ast 0}D^{\ast +} $, $B_{s}^{0}D_{s}^{+}$, and $B_{s}^{\ast 0}D_{s}^{\ast
+}$ are studied to find the width of the $\widetilde{H}_{\mathrm{bc}}$
state. The widths of these processes are calculated using QCD three-point
sum rule approach, which  require estimation of the strong couplings at
hybrid-meson-meson vertices. The results $(50.8 \pm 9.8)~\mathrm{MeV}$ and $%
(184.3\pm 22.8)~\mathrm{MeV}$ for the full widths of the hybrid mesons $H_{%
\mathrm{bc}}$ and $\widetilde{H}_{\mathrm{bc}}$ characterize them as
relatively narrow and broad structures, respectively.
\end{abstract}

\maketitle


\section{Introduction}

\label{sec:Intro}

Due to recent experimental achievements,  the exotic hadrons became objects of
intensive theoretical investigations. Actually, mesons and baryons which can
not be included into the ordinary $q\overline{q}^{\prime }$ and $qq^{\prime
}q^{\prime \prime }$ scheme already attracted interests of researches. First
predictions for the masses, production and decay mechanisms of these
hypothetical particles were obtained decades before the experimental
discoveries.

The hybrid mesons, in other words, mesons containing apart from the valence
quark-antiquarks also valence gluon(s) form a class of such exotic states.
Existence of hybrid mesons and baryons does not contradict to fundamental
principles of QCD and quark-gluon model. Investigations of hybrid hadrons
were started in publications Ref. \cite{Jaffe:1975fd,Horn:1977rq} and
continue till now supported by new ideas, models, and computational methods
\cite{Meyer:2015eta}.

Experimentally observed resonances which may be interpreted as hybrid mesons
are light vector particles $\pi _{1}(1600)$ and $\pi _{1}(2015)$ with
spin-parities $J^{\mathrm{PC}}=$ ${1^{-+}}$ \cite{PDG:2024}. They were
discovered by the $\mathrm{E892}$ collaboration in the reaction $\pi ^{-}p$
and reported in different publications \cite%
{E852:2001ikk,E852:2004gpn,E852:2004rfa}. It is worth noting that the
resonances ${1^{-+}}$ were also observed by other experimental groups. Thus,
the isoscalar particle $\eta _{1}(1855)$ bearing quantum numbers $J^{\mathrm{%
PC}}=$ ${1^{-+}}$ was fixed quite recently in the radiative decay $J/\psi
\rightarrow \gamma \eta _{1}(1855)\rightarrow \gamma \eta \eta ^{\prime }$
\cite{BESIII:2022riz}.

There are structures seen in different decays that may be considered as the
hybrid charmonium states. For instance, the resonances $\psi (4390)$ and $%
\psi (4660)$ with $J^{\mathrm{PC}}=$ ${1^{--}}$ maybe are excited spin
singlet hybrid charmonia \cite{Brambilla:2022hhi}. The latter can be
interpreted also as the ground-level vector tetraquark $c\overline{c}s%
\overline{s}$ \cite{Sundu:2018toi}. Some of the resonances in the
bottomonium sector may be interpreted as exotic hybrid mesons as well. List
of resonances which are candidates to hybrid quarkonia was presented in
Ref.\ \cite{Brambilla:2022hhi}.

Theoretical explorations of the hybrid mesons encompass considerably wider
range of objects with different spin-parities. Light or heavy-light hybrids
are, naturally, on focus of interesting studies \cite%
{Barsbay:2022gtu,Barsbay:2024vjt} (and references therein). The hybrid
bottomonia $\overline{b}gb$, charmonia $\overline{c}gc$, as well as the
mesons $\overline{b}gc$ were and remain on agenda of high energy physics
\cite%
{Govaerts:1984hc,Govaerts:1985fx,Page:1998gz,Zhu:1998sv,Qiao:2010zh,Harnett:2012gs, HadronSpectrum:2012gic,Chen:2013zia,Chen:2013eha,Cheung:2016bym,Palameta:2018yce,Miyamoto:2019oin, Brambilla:2018pyn,Ryan:2020iog,TarrusCastella:2021pld,Woss:2020ayi,Soto:2023lbh,Bruschini:2023tmm,Wang:2024hvp}%
. It is worth to note that in works devoted to analysis of hybrid
structures, authors used various methods and model to calculate their
parameters. Among widely applied tools of theoretical studies are various
quark-gluon models, the lattice simulations, and QCD sum rule (SR) method.
Predictions for the masses of hybrid structures made in these papers, as
usual, differ from each other. Therefore, studies carried out with higher
accuracy can shed light on existing problems and help to clarify the situation
with heavy hybrid states.

The hybrid quarkonia $\overline{b}gb$, $\overline{c}gc$ with spin-parities $%
J^{\mathrm{PC}}=0^{(\pm )(\pm )}$ and $1^{(\pm )(\pm )}$ and mesons $%
\overline{b}gc$ with $J^{\mathrm{P}}=0^{\pm }$, $1^{\pm }$ were also
explored in our paper \cite{Alaakol:2024zyh}. In this article, we calculated
the masses and current couplings of these states. The tensor charmonia $%
\overline{c}gc$ with $J^{\mathrm{PC}}=2^{-+}$ and $2^{++}$ were considered
in Ref.\ \cite{Agaev:2025llz}, in which their masses and
widths  were determined.

In the present work, we consider the tensor hybrid mesons $H_{\mathrm{bc}}$
and $\widetilde{H}_{\mathrm{bc}}$ with content $\overline{b}gc$ and quantum
numbers $J^{\mathrm{P}}=$ ${2^{-}}$ and ${2^{+}}$, respectively. We are
going to calculate the masses of these mesons. For these purposes, we apply
the QCD two-point SR method by taking into account nonperturbative terms up
to dimension-$12$. We compute also the full widths of these states. The
width of $H_{\mathrm{bc}}$ is found by exploring the processes $H_{\mathrm{bc%
}}\rightarrow B^{\ast 0}D^{+}$ and $B^{\ast +}D^{0}$. Decays $\widetilde{H}_{%
\mathrm{bc}}\rightarrow B^{+}D^{0}$, $B^{0}D^{+}$, $B^{\ast +}D^{\ast 0}$, $%
B^{\ast 0}D^{\ast +}$, $B_{s}^{0}D_{s}^{+}$, and $B_{s}^{\ast 0}D_{s}^{\ast
+}$ are allowed modes of the hybrid meson $\widetilde{H}_{\mathrm{bc}}$: The
full width of $\widetilde{H}_{\mathrm{bc}}$ is saturated by these decay
channels. The partial widths of all  these decays are evaluated using QCD
three-point SR approach: It enables us to evaluate couplings at the various
hybrid-meson-meson vertices.

This work is structured in the following way. In Sec.\ \ref{sec:Hmeson}, we
compute the spectral parameters of the tensor mesons $\overline{b}gc$ with $%
J^{\mathrm{P}}=2^{-}$ and $\ 2^{+}$. The decays of $H_{\mathrm{bc}}$ are
considered in Section \ref{sec:Hbcdecays}. In Sec.\ \ref{sec:TildeHbcWidth}
we estimate the full width of $\widetilde{H}_{\mathrm{bc}}$. In the last
section we sum up our results.


\section{Masses and current couplings of $H_{\mathrm{bc}}$ and $\widetilde{H}%
_{\mathrm{bc}}$}

\label{sec:Hmeson}

Here, we study the tensor hybrid mesons $H_{\mathrm{bc}}$ and $\widetilde{H}%
_{\mathrm{bc}}$ composed of $\overline{b}gc$ quark-gluons and spin-parities $%
J^{\mathrm{P}}=2^{-}$ and $2^{+}$, respectively. Investigations are
performed in the context of QCD SR method \cite%
{Shifman:1978bx,Shifman:1978by} suggested to study properties of ordinary
hadrons. But this approach can also be used to explore the exotic hadrons as
well \cite{Nielsen:2009uh,Albuquerque:2018jkn,Agaev:2020zad}. Note that this  approach was used to explore the hybrid quarkonia in first years of its
invention \cite{Govaerts:1984hc,Govaerts:1985fx}.

The sum rules for the spectral parameters $m$ and $\Lambda $ of the meson $%
H_{\mathrm{bc}}=\overline{b}gc$ can be extracted from analysis of the
correlation function
\begin{equation}
\Pi _{\mu \nu \mu ^{\prime }\nu ^{\prime }}(p)=i\int d^{4}xe^{ipx}\langle 0|%
\mathcal{T}\{J_{\mu \nu }(x)J_{\mu ^{\prime }\nu ^{\prime }}^{\dag
}(0)\}|0\rangle .  \label{eq:CF1}
\end{equation}%
Here, $J_{\mu \nu }(x)$ is the interpolating current for the structure under
analysis, and $\mathcal{T}$ \ is a time-ordered product of two currents.

The tensor hybrid meson $H_{\mathrm{bc}}$ can be interpolated by the current
\begin{equation}
J_{\mu \nu }(x)=g_{s}\overline{b}_{a}(x)\sigma _{\mu }^{\alpha }\gamma _{5}%
\frac{{\lambda }_{ab}^{n}}{2}G_{\alpha \nu }^{n}(x)c_{b}(x).  \label{eq:C1}
\end{equation}%
In the case of $\widetilde{H}_{\mathrm{bc}}$ the current $\widetilde{J}_{\mu
\nu }(x)$ is given by the formula
\begin{equation}
\widetilde{J}_{\mu \nu }(x)=g_{s}\overline{b}_{a}(x)\sigma _{\mu }^{\alpha
}\gamma _{5}\frac{{\lambda }_{ab}^{n}}{2}\widetilde{G}_{\alpha \nu
}^{n}(x)c_{b}(x),  \label{eq:C2}
\end{equation}%
where $a$ and $b$ are the color indices, and $g_{s}$ is the QCD strong
coupling constant. In Eqs.\ (\ref{eq:C1}) and (\ref{eq:C2}), $b_{a}(x)$ and $%
c_{a}(x)$ are quark fields, whereas $G_{\mu \nu }^{n}(x)$ is a strength
tensor of the gluon field. The relevant dual tensor is denoted by $%
\widetilde{G}_{\mu \nu }^{n}(x)=\varepsilon _{\mu \nu \alpha \beta
}G^{n\alpha \beta }(x)/2$. The Gell-Mann matrices are labeled as ${\lambda }%
^{n}$,  $n=1,2,..8$.

We begin from analysis of the tensor hybrid meson $H_{\mathrm{bc}}$ with the
current $J_{\mu \nu }(x)$. We write $\Pi _{\mu \nu \mu ^{\prime }\nu
^{\prime }}(p)$ using the mass and current coupling of the hybrid state
\begin{eqnarray}
&&\Pi _{\mu \nu \mu ^{\prime }\nu ^{\prime }}^{\mathrm{Phys}}(p)=\frac{%
\langle 0|J_{\mu \nu }|H_{\mathrm{bc}}(p,\epsilon )\rangle \langle H_{%
\mathrm{bc}}(p,\epsilon )|J_{\mu ^{\prime }\nu ^{\prime }}^{\dag }|0\rangle
}{m^{2}-p^{2}}  \notag \\
&&+\cdots ,  \label{eq:CF2}
\end{eqnarray}%
where $\epsilon =\epsilon _{\mu \nu }^{(\lambda )}(p)$ is the polarization
tensor of $H_{\mathrm{bc}}$. Here, the ground-level term is presented
explicitly, while contributions of the higher resonances and continuum
states are shown by the dots. To continue,  we also introduce the matrix
element
\begin{equation}
\langle 0|J_{\mu \nu }|H_{\mathrm{bc}}(p,\epsilon )\rangle =\Lambda \epsilon
_{\mu \nu }^{(\lambda )}(p).  \label{eq:M1}
\end{equation}%
To determine $\Pi _{\mu \nu \mu ^{\prime }\nu ^{\prime }}^{\mathrm{Phys}}(p)$%
, we insert Eq.\ (\ref{eq:M1}) into the correlation function Eq.\ (\ref%
{eq:CF2}) and carry out summation over polarization tensor $\epsilon _{\mu
\nu }^{(\lambda )}(p)$. To this end, we apply the identity
\begin{eqnarray}
\sum\limits_{\lambda }\epsilon _{\mu \nu }^{(\lambda )}(p)\epsilon _{\mu
^{\prime }\nu ^{\prime }}^{\ast (\lambda )}(p) &=&\frac{1}{2}(\widetilde{g}%
_{\mu \mu ^{\prime }}\widetilde{g}_{\nu \nu ^{\prime }}+\widetilde{g}_{\mu
\nu ^{\prime }}\widetilde{g}_{\nu \mu ^{\prime }})  \notag \\
&&-\frac{1}{3}\widetilde{g}_{\mu \nu }\widetilde{g}_{\mu ^{\prime }\nu
^{\prime }},  \label{eq:F1}
\end{eqnarray}%
where%
\begin{equation}
\widetilde{g}_{\mu \nu }=-g_{\mu \nu }+\frac{p_{\mu }p_{\nu }}{p^{2}}.
\label{eq:F2}
\end{equation}%
Our computations yield
\begin{eqnarray}
\Pi _{\mu \nu \mu ^{\prime }\nu ^{\prime }}^{\mathrm{Phys}}(p) &=&\frac{%
\Lambda ^{2}}{m^{2}-p^{2}}\left\{ \frac{1}{2}\left( g_{\mu \mu ^{\prime
}}g_{\nu \nu ^{\prime }}+g_{\mu \nu ^{\prime }}g_{\nu \mu ^{\prime }}\right)
\right.  \notag \\
&&\left. +\text{other structures}\right\} +\cdots .  \label{eq:Phys2}
\end{eqnarray}%
Application of Eqs.\ (\ref{eq:F1}) and (\ref{eq:F2}) gives rise to numerous
terms in $\Pi _{\mu \nu \mu ^{\prime }\nu ^{\prime }}^{\mathrm{Phys}}(p)$.
These components receive contributions from particles of the different
spins. The term proportional to $(g_{\mu \mu ^{\prime }}g_{\nu \nu ^{\prime
}}+g_{\mu \nu ^{\prime }}g_{\nu \mu ^{\prime }})$ is formed due to
contribution of only spin-$2$ particle. Therefore, in our investigations we
consider this Lorentz structure and corresponding invariant amplitude $\Pi ^{%
\mathrm{Phys}}(p^{2})$.

At the next stage of calculations, we compute the correlator $\Pi _{\mu \nu
\mu ^{\prime }\nu ^{\prime }}(p)$ using the operator product expansion ($%
\mathrm{OPE}$). For these purposes, we employ $J_{\mu \nu }(x)$ in Eq.\ (\ref%
{eq:CF1}) and contract corresponding quark and gluon fields. Then, we get
\begin{eqnarray}
&&\Pi _{\mu \nu \mu ^{\prime }\nu ^{\prime }}^{\mathrm{OPE}}(p)=\frac{%
ig_{s}^{2}}{4}\int d^{4}xe^{ipx}{\lambda }_{ab}^{n}{\lambda }_{a^{\prime
}b^{\prime }}^{m}\langle 0|G_{\alpha \nu }^{n}(x)G_{\alpha ^{\prime }\nu
^{\prime }}^{m}(0)|0\rangle  \notag \\
&&\times \mathrm{Tr}\left[ \sigma _{\mu }^{\alpha }\gamma
_{5}S_{c}^{bb^{\prime }}(x)\gamma _{5}\sigma _{\mu ^{\prime }}^{\alpha
^{\prime }}S_{b}^{a^{\prime }a}(-x)\right] ,  \label{eq:OPE1}
\end{eqnarray}%
where $S_{c(b)}^{ab}(x)$ are the heavy quark propagators presented in Ref.\ \cite%
{Agaev:2020zad}.

The correlation function $\Pi _{\mu \nu \mu ^{\prime }\nu ^{\prime }}^{%
\mathrm{OPE}}(p)$ depends on two main factors: These are the trace term and
gluon's vacuum matrix element. The former contains the $b$ and $c$ quark
propagators which have the perturbative and nonperturbative components. The
nonperturbative components in $S_{c(b)}^{ab}(x)$ include terms proportional
to $g_{s}G_{ab}^{\alpha \beta }$, $g_{s}^{2}G^{2}$ and $g_{s}^{3}G^{3}$. The
last two ones placed between the vacuum states generate gluon condensates $%
\langle g_{s}^{2}G^{2}\rangle $ and $\langle g_{s}^{3}G^{3}\rangle $. A term
$\sim g_{s}G_{ab}^{\alpha \beta }$ in $S_{c}^{ab}(x)$ together with a
similar term in the second propagator gives rise to additional conrtibutions
$\langle g_{s}^{2}G^{2}\rangle $, which are taken into account as well.

The second factor in Eq.\ (\ref{eq:OPE1}) is the matrix element $\langle
0|G_{\alpha \nu }^{n}(x)G_{\alpha ^{\prime }\nu ^{\prime }}^{m}(0)|0\rangle $%
. Its treatment is twofold: we replace it by the condensate $\langle
g_{s}^{2}G^{2}\rangle $ and utilize
\begin{eqnarray}
&&\langle 0|g_{s}^{2}G_{\alpha \nu }^{n}(x)G_{\alpha ^{\prime }\nu ^{\prime
}}^{m}(0)|0\rangle =\frac{\langle g_{s}^{2}G^{2}\rangle }{96}\delta ^{nm}
\notag \\
&&\times \left[ g_{\alpha \alpha ^{\prime }}g_{\nu \nu ^{\prime }}-g_{\alpha
\nu ^{\prime }}g_{\alpha ^{\prime }\nu }\right] ,  \label{eq:GluonME1}
\end{eqnarray}%
which is the first term in the expansion at $x=0$. Alternatively, we employ
the full gluon propagator in the $x$-space
\begin{eqnarray}
&&\langle 0|G_{\alpha \nu }^{n}(x)G_{\alpha ^{\prime }\nu ^{\prime
}}^{m}(0)|0\rangle =\frac{\delta ^{nm}}{2\pi ^{2}x^{4}}\left[ g_{\nu \nu
^{\prime }}\left( g^{\alpha \alpha ^{\prime }}-\frac{4x_{\alpha }x_{\alpha
^{\prime }}}{x^{2}}\right) \right.  \notag \\
&&\left. +(\nu ,\nu ^{\prime })\leftrightarrow (\alpha ,\alpha ^{\prime
})-\nu \leftrightarrow \alpha -\nu ^{\prime }\leftrightarrow \alpha ^{\prime
}\right] .  \label{eq:GluonME2}
\end{eqnarray}%
Terms generated by Eq.\ (\ref{eq:GluonME1}) leads to gluon-QCD vacuum
diagrams, whereas the alternative treatment gives rise to diagrams with full
valence gluon propagator.

Having fixed the structure $(g_{\mu \mu ^{\prime }}g_{\nu \nu ^{\prime
}}+g_{\mu \nu ^{\prime }}g_{\nu \mu ^{\prime }})$ from $\Pi _{\mu \nu \mu
^{\prime }\nu ^{\prime }}^{\mathrm{OPE}}(p)$ and denoted the relevant
amplitude by $\Pi ^{\mathrm{OPE}}(p^{2})$, one can derive SRs for the
parameters of the meson $H_{\mathrm{bc}}$ which read
\begin{equation}
m^{2}=\frac{\Pi ^{\prime }(M^{2},s_{0})}{\Pi (M^{2},s_{0})},  \label{eq:Mass}
\end{equation}%
and
\begin{equation}
\Lambda ^{2}=e^{m^{2}/M^{2}}\Pi (M^{2},s_{0}).  \label{eq:Coupl}
\end{equation}%
In Eq.\ (\ref{eq:Mass}), we also use the short-hand notation $\Pi ^{\prime
}(M^{2},s_{0})=d\Pi (M^{2},s_{0})/d(-1/M^{2})$. Here,
\begin{equation}
\Pi (M^{2},s_{0})=\int_{(m_{b}+m_{c})^{2}}^{s_{0}}ds\rho ^{\mathrm{OPE}%
}(s)e^{-s/M^{2}},  \label{eq:CorrF}
\end{equation}%
where $s_{0}$ is the continuum subtraction parameter. The spectral density $%
\rho ^{\mathrm{OPE}}(s)$ contains the perturbative $\rho ^{\mathrm{pert.}%
}(s) $ and nonperturbative $\rho ^{\mathrm{DimN}}(s)$ terms ($\mathrm{%
N=4,6,8,10,12}$). We refrain from presenting their explicit expressions here.

We need to specify the input parameters in Eqs.\ (\ref{eq:Mass}) and (\ref%
{eq:Coupl}) to perform numerical computations. Some of them are universal
quantities and do not depend on the  problem under consideration. The masses of
$b$ and $c$ quarks and gluon vacuum condensates $\langle \alpha
_{s}G^{2}/\pi \rangle $ and $\langle g_{s}^{3}G^{3}\rangle $ are such
parameters. In present work, we use the following values
\begin{eqnarray}
&&m_{c}(\mu )=(1.27\pm 0.02)~\mathrm{GeV},\ m_{b}(\mu )=4.18_{-0.02}^{+0.03}~%
\mathrm{GeV},  \notag \\
&&\langle \alpha _{s}G^{2}/\pi \rangle =(0.012\pm 0.004)~\mathrm{GeV}^{4},
\notag \\
&&\langle g_{s}^{3}G^{3}\rangle =(0.57\pm 0.29)~\mathrm{GeV}^{6}.
\label{eq:GluonCond}
\end{eqnarray}

The $m_{c}$ and $m_{b}$ are the $\overline{\mathrm{MS}}$ scheme running
masses at $\mu =m_{c}$ and $\mu =m_{b}$ \cite{PDG:2024}, respectively. The
gluon condensates were found from exploration of hadronic processes \cite%
{Shifman:1978bx,Shifman:1978by,Narison:2015nxh}.

The parameters $M^{2}$ and $s_{0}$ that enter  Eqs.\ (\ref{eq:Mass}) and (%
\ref{eq:Coupl}) are specific for  each process and have to be chosen to
satisfy standard constraints of SR investigations. Stated differently, these
parameters should lead to the dominance of the pole contribution ($\mathrm{PC%
}$) in the correlators. The convergence of the $\mathrm{OPE}$ and stability
of results against variations of $M^{2}$ are important restrictions as well.
To control these parameters of the SR analysis, we utilize
\begin{equation}
\mathrm{PC}=\frac{\Pi (M^{2},s_{0})}{\Pi (M^{2},\infty )},  \label{eq:PC}
\end{equation}%
and%
\begin{equation}
R(M^{2})=\frac{\Pi ^{\mathrm{DimN}}(M^{2},s_{0})}{\Pi (M^{2},s_{0})},
\label{eq:Conv}
\end{equation}%
where $\Pi ^{\mathrm{DimN}}(M^{2},s_{0})=\sum_{\mathrm{N}=8,10,12}\Pi ^{%
\mathrm{DimN}}$ is a sum of last three terms in $\mathrm{OPE}$ which are
proportional to $\langle g_{s}^{2}G^{2}\rangle ^{2}$, $\langle
g_{s}^{2}G^{2}\rangle \langle g_{s}^{3}G^{3}\rangle $ and $\langle
g_{s}^{3}G^{3}\rangle ^{2}$, respectively.

\begin{figure}[h]
\includegraphics[width=8.5cm]{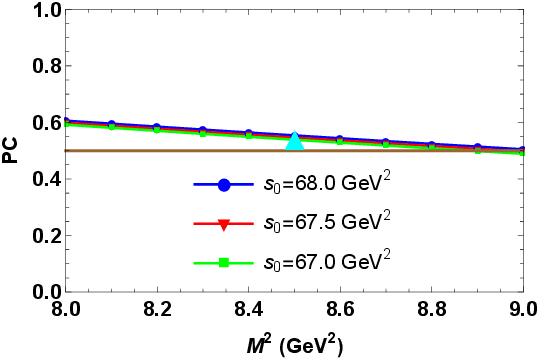}
\caption{Pole contribution $\mathrm{PC}$ as a function of $M^{2}$ at fixed $%
s_{0}$. The triangle marks the point $M^{2}=8.5~\mathrm{GeV}^{2}$ and $%
s_{0}=67.5~\mathrm{GeV}^{2}$. }
\label{fig:PC}
\end{figure}

Numerical calculations show that in the case of $H_{\mathrm{bc}}$ the
regions
\begin{equation}
M^{2}\in \lbrack 8,9]~\mathrm{GeV}^{2},\ s_{0}\in \lbrack 67,68]~\mathrm{GeV}%
^{2},  \label{eq:Wind1}
\end{equation}%
meet constraints of SR analysis. Thus, at $M^{2}=9~\mathrm{GeV}^{2}$ and $%
M^{2}=8~\mathrm{GeV}^{2}$ on the average in $s_{0}$ the pole contribution is
$\mathrm{PC}\approx 0.5$ and $\mathrm{PC}$ $\approx 0.6$, respectively. At $%
M^{2}=8~\mathrm{GeV}^{2}$ the parameter $R(M^{2})$ is positive and less than
$0.02$. The size of the region for the Borel parameter is determined by the
restriction $\mathrm{PC}\geq 0.5$ imposed on the pole contribution. The $%
\mathrm{PC}$ changes depending on $M^{2}$ which is vizualized in Fig.\ \ref%
{fig:PC}.

The mass $m$ and current coupling $\Lambda $ are found as their average
values over the regions Eq.\ (\ref{eq:Wind1}). They amount to
\begin{eqnarray}
m &=&(7.214\pm 0.075)~\mathrm{GeV},\   \notag \\
\Lambda &=&(0.54\pm 0.04)~\mathrm{GeV}^{4},  \label{eq:Result1}
\end{eqnarray}%
respectively. These results coincide with the SR predictions at $M^{2}=8.5~%
\mathrm{GeV}^{2}$ and $s_{0}=67.5~\mathrm{GeV}^{2}$, where the pole
contribution is $\mathrm{PC}\approx 0.55$.

The main source of uncertainties in Eq.\ (\ref{eq:Result1}) are the
parameters $M^{2}$, $s_{0}$ and $\langle \alpha _{s}G^{2}/\pi \rangle $.
Uncertainties of the condensate $\langle g_{s}^{3}G^{3}\rangle $ lead to
corrections which are very small. In this work, for $\langle
g_{s}^{3}G^{3}\rangle $ we use its central value. The dependence of $m$ on
the Borel and continuum subtraction parameters is depicted in Fig.\ \ref%
{fig:Mass1}.

\begin{widetext}

\begin{figure}[h]
\begin{center}
\includegraphics[totalheight=6cm,width=8cm]{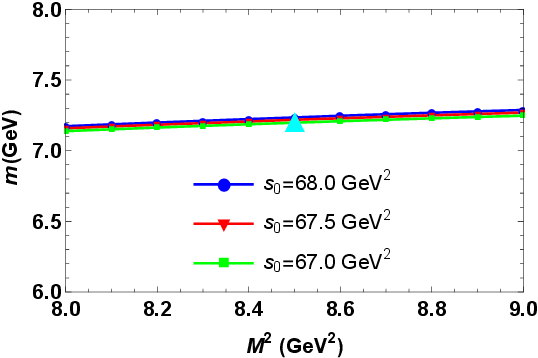} %
\includegraphics[totalheight=6cm,width=8cm]{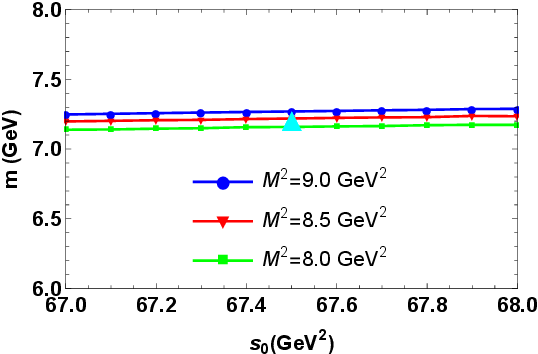}
\end{center}
\caption{Dependence of the mass $m$ of the hybrid meson $H_{\mathrm{bc}}$ on the
Borel $M^{2}$ (left panel), and continuum threshold $s_{0}$ parameters
(right panel).}
\label{fig:Mass1}
\end{figure}

\begin{figure}[h]
\begin{center}
\includegraphics[totalheight=6cm,width=8cm]{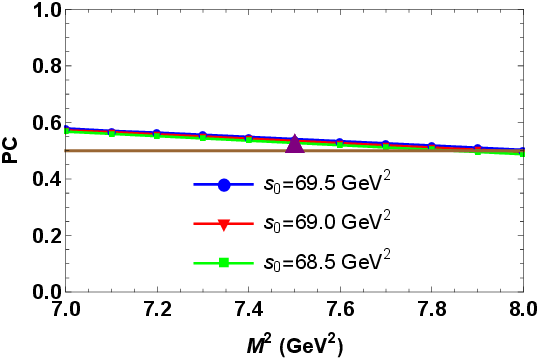} %
\includegraphics[totalheight=6cm,width=8cm]{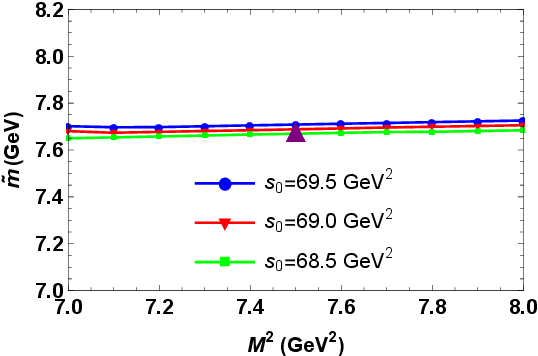}
\end{center}
\caption{$\mathrm{PC}$ (left) and $\widetilde{m}$ (right) of the hybrid meson $\widetilde{H}_{\mathrm{bc}}$ as functions of the Borel parameter. The red triangle fixes the point $M^2=7.5~\mathrm{GeV}^2$ and
$s_0=69~\mathrm{GeV}^2$.}
\label{fig:Mass2}
\end{figure}

\end{widetext}

The hybrid meson $\widetilde{H}_{\mathrm{bc}}$ with the spin-parity $J^{%
\mathrm{P}}=2^{+}$ is studied by the same way. Thus, the correlator $%
\widetilde{\Pi }_{\mu \nu \mu ^{\prime }\nu ^{\prime }}^{\mathrm{OPE}}(p)$
can be obtained from Eq.\ (\ref{eq:OPE1}) by replacement $G\rightarrow
\widetilde{G}$. The working windows for the parameters $M^{2}$ and $s_{0}$
are
\begin{equation}
M^{2}\in \lbrack 7,8]~\mathrm{GeV}^{2},\ s_{0}\in \lbrack 68.5,69.5]~\mathrm{%
GeV}^{2},  \label{eq:Wind1A}
\end{equation}%
where the pole contribution changes within the limits $0.50\leq \mathrm{PC}%
\leq 0.57$. For the spectroscopic parameters $\widetilde{m}$ and $\widetilde{%
\Lambda }$ of the hybrid meson $\widetilde{H}_{\mathrm{bc}}$, one obtains
\begin{eqnarray}
\widetilde{m} &=&(7.685\pm 0.040)~\mathrm{GeV},\ \   \notag \\
\widetilde{\Lambda } &=&(0.67\pm 0.03)~\mathrm{GeV}^{4},
\end{eqnarray}%
which are effectively equal to SR results at the point $M^{2}=7.5~\mathrm{GeV%
}^{2}$ and $s_{0}=69~\mathrm{GeV}^{2}$, where $\mathrm{PC}\approx 0.54$. The
pole contribution and the mass $\widetilde{m}$ as functions of the parameter
$M^{2}$ are plotted in Fig.\ \ref{fig:Mass2}.

The tensor hybrid mesons $H_{\mathrm{bc}}$ and $\widetilde{H}_{\mathrm{bc}}$
in the context of the SR method were also studied in Ref.\ \cite%
{Chen:2013eha}. The masses of these exotic states were found there equal to $%
(7.15\pm ~0.13)~\mathrm{GeV}$ and $(7.67\pm ~0.12)~\mathrm{GeV}$,
respectively. These results are very close to our predictions: The mass of
the meson $J^{\mathrm{P}}=2^{+}$ almost coincides with $\widetilde{m}$.

The next problem that should be addressed in this article is computation of
the widths of the hybrid structures $H_{\mathrm{bc}}$ and $\widetilde{H}_{%
\mathrm{bc}}$. Knowledge about masses of these mesons gained in this work
enables us to determine their kinematically allowed decay channels.


\section{Decays of the meson $H_{\mathrm{bc}}$}

\label{sec:Hbcdecays}

This section is devoted to exploration of decays of the hybrid meson $H_{%
\mathrm{bc}}$. The two-body decays $H_{\mathrm{bc}}\rightarrow B^{\ast
0}D^{+}$ and $B^{\ast +}D^{0}$ are possible modes of $H_{\mathrm{bc}}$.
Indeed, the central value of $H_{\mathrm{bc}}$ meson's mass exceeds
thresholds $7.195~\mathrm{GeV}$ for these processes. Because the mass gaps
between $m$ and thresholds are small, at the low values of $m$ in Eq.\ (\ref%
{eq:Result1}) $H_{\mathrm{bc}}$ can not decay to these mesons. Therefore, in
the analysis we will use the central value $7.214~\mathrm{GeV}$ for the mass
of the hybrid state $H_{\mathrm{bc}}$.

The partial widths of these processes depend on the strong couplings $g_{i}$
at the $H_{\mathrm{bc}}$-meson-meson vertices. To determine $g_{i}$, one
uses the three-point SR approach that enables one to find the form factors $%
g_{i}(q^{2})$ and fix at $q^{2}=m_{D}^{2}$ the corresponding couplings. It
is worth noting that in the current paper we accept the approximation $%
m_{u}=m_{d}=0$, therefore after replacement $d\rightarrow u$ the first decay
turns to the second one. Because, we employ the same decay constants for the
charged and neutral $B^{\ast }$ and $D$ mesons, as well as neglect the small
mass differences between them, the width of the decay $H_{\mathrm{bc}%
}\rightarrow B^{\ast +}D^{0}$ is equal to that of the process $H_{\mathrm{bc}%
}\rightarrow $ $B^{\ast 0}D^{+}$ .

The form factor $g_{1}(q^{2})$ which characterize the strong interaction of
particles at the vertex $H_{\mathrm{bc}}B^{\ast 0}D^{+}$ is obtained from
analysis of the correlation function
\begin{eqnarray}
\Pi _{\mu \alpha \beta }(p,p^{\prime }) &=&i^{2}\int
d^{4}xd^{4}ye^{ip^{\prime }x}e^{iqy}\langle 0|\mathcal{T}\{J_{\mu }^{B^{\ast
0}}(x)  \notag \\
&&\times J^{D^{+}}(y)J_{\alpha \beta }^{\dagger }(0)\}|0\rangle ,
\label{eq:CF1a}
\end{eqnarray}%
where $J_{\mu }^{B^{\ast 0}}(x)$ and $J^{D^{+}}(x)$ are currents that
interpolate the mesons $B^{\ast 0}$ and $D^{+}$, respectively. They are
introduced in the form%
\begin{equation}
J_{\mu }^{B^{\ast 0}}(x)=\overline{b}_{i}(x)\gamma _{\mu }d_{i}(x),\
J^{D^{+}}(x)=\overline{d}_{j}(x)i\gamma _{5}c_{j}(x),  \label{eq:CDmes}
\end{equation}%
where $i$ and $j$ are the color indices.

To find the physical side of the sum rule $\Pi _{\mu \alpha \beta }^{\mathrm{%
Phys}}(p,p^{\prime })$, we rewrite Eq.\ (\ref{eq:CF1a}) using the parameters
of the particles. By keeping, in the formula,  the  contribution of the
ground-level particles, we recast the correlator $\Pi _{\mu \alpha \beta
}(p,p^{\prime })$ into the form
\begin{eqnarray}
&&\Pi _{\mu \alpha \beta }^{\mathrm{Phys}}(p,p^{\prime })=\frac{\langle
0|J_{\mu }^{B^{\ast 0}}|B^{\ast 0}(p^{\prime },\varepsilon )\rangle }{%
p^{\prime 2}-m_{D^{\ast }}^{2}}\frac{\langle 0|J^{D^{+}}|D^{+}(q)\rangle }{%
q^{2}-m_{D}^{2}}  \notag \\
&&\times \langle B^{\ast 0}(p^{\prime },\varepsilon )D^{+}(q)|H_{\mathrm{bc}%
}(p,\epsilon )\rangle \frac{\langle H_{\mathrm{bc}}(p,\epsilon )|J_{\alpha
\beta }^{\dagger }|0\rangle }{p^{2}-m^{2}}  \notag \\
&&+\cdots ,  \label{eq:TP1}
\end{eqnarray}%
where $m_{B^{\ast 0}}=(5324.75\pm 0.20)~\mathrm{MeV}$ and $m_{D}=(1869.5\pm
0.4)~\mathrm{MeV}$ are the masses of the $B^{\ast 0}$ and $D^{+}$ mesons
\cite{PDG:2024}, respectively.

We also use the following matrix elements
\begin{eqnarray}
0|J_{\mu }^{B^{\ast 0}}|B^{\ast 0}(p^{\prime },\varepsilon )\rangle
&=&f_{B^{\ast }}m_{B^{\ast 0}}\varepsilon _{\mu }(p^{\prime }),  \notag \\
\langle 0|J^{D^{+}}|D^{+}(q)\rangle &=&\frac{f_{D}m_{D}^{2}}{m_{c}}.
\label{eq:ME1}
\end{eqnarray}%
In Eq.\ (\ref{eq:ME1}) $f_{B^{\ast }}=(210\pm 6)~\mathrm{MeV}$ and $%
f_{D}=(212.0\pm 0.7)~\mathrm{MeV}$ are decay constants of $B^{\ast 0}$ and $%
D^{+}$. Above, the polarization vector of the meson $B^{\ast }{}^{0}$ is
denoted by $\varepsilon _{\mu }(p^{\prime })$. The vertex matrix element $%
\langle B^{\ast 0}(p^{\prime },\varepsilon )D^{+}(q)|H_{\mathrm{bc}%
}(p,\epsilon )\rangle $ is introduced as
\begin{eqnarray}
\langle B^{\ast 0}(p^{\prime },\varepsilon )D^{+}(q)|H_{\mathrm{bc}%
}(p,\epsilon )\rangle &=&g_{1}(q^{2})\epsilon ^{\rho \sigma }(p)q_{\rho
}\varepsilon _{\sigma }(p^{\prime }).  \notag \\
&&
\end{eqnarray}%
Then, for $\Pi _{\mu \alpha \beta }^{\mathrm{Phys}}(p,p^{\prime })$ one
obtains
\begin{eqnarray}
&&\Pi _{\mu \alpha \beta }^{\mathrm{Phys}}(p,p^{\prime })=g_{1}(q^{2})\frac{%
\Lambda f_{B^{\ast }}m_{B^{\ast 0}}f_{D}m_{D}^{2}}{m_{c}\left(
p^{2}-m^{2}\right) (p^{\prime 2}-m_{B^{\ast 0}}^{2})}  \notag \\
&&\times \frac{1}{(q^{2}-m_{D}^{2})}\left[ \frac{1}{2}g_{\alpha \beta
}p_{\mu }^{\prime }-\frac{(m^{2}+m_{B^{\ast 0}}^{2}-q^{2})^{2}}{%
12m^{2}m_{B^{\ast 0}}^{2}}g_{\mu \alpha }p_{\beta }^{\prime }\right.  \notag
\\
&&\left. +\text{other structures}\right] +\cdots .  \label{eq:PhysSide}
\end{eqnarray}

The correlator $\Pi _{\mu \alpha \beta }(p,p^{\prime })$ computed in terms
of the quark propagators and the gluon field is%
\begin{eqnarray}
&&\Pi _{\mu \alpha \beta }^{\mathrm{OPE}}(p,p^{\prime })=-i\int
d^{4}xd^{4}ye^{ip^{\prime }x}e^{iqy}g_{s}\frac{{\lambda }_{ab}^{n}}{2}%
\widetilde{G}_{\rho \alpha }^{n}(0)  \notag \\
&&\times \mathrm{Tr}\left[ \gamma _{\beta }S_{d}^{ij}(x-y)\gamma
_{5}S_{c}^{jb}(y)\gamma _{5}\sigma _{\mu }^{\rho }S_{b}^{ai}(-x)\right] ,
\label{eq:CF3}
\end{eqnarray}%
where $S_{d}^{ij}(x-y)$ is the light quark propagator \cite{Agaev:2020zad}.
The sum rule for $g_{1}(q^{2})$ is obtained by employing the invariant
amplitudes $\Pi _{1}^{\mathrm{Phys}}(p^{2},p^{\prime 2},q^{2})$ and $\Pi
_{1}^{\mathrm{OPE}}(p^{2},p^{\prime 2},q^{2})$ which correspond to
structures $g_{\alpha \beta }p_{\mu }^{\prime }$ in the correlation
functions. After usual manipulations the sum rule for $g_{1}(q^{2})$ reads
\begin{equation}
g_{1}(q^{2})=\frac{2m_{c}(q^{2}-m_{D}^{2})}{\Lambda f_{B^{\ast }}m_{B^{\ast
}}f_{D}m_{D}^{2}}e^{m^{2}/M_{1}^{2}}e^{m_{B^{\ast }}^{2}/M_{2}^{2}}\Pi _{1}(%
\mathbf{M}^{2},\mathbf{s}_{0},q^{2}).  \label{eq:SRG}
\end{equation}%
In Eq.\ (\ref{eq:SRG}), $\Pi _{1}(\mathbf{M}^{2},\mathbf{s}_{0},q^{2})$ is
the function $\Pi _{1}^{\mathrm{OPE}}(p^{2},p^{\prime 2},q^{2})$ after the
Borel transformations and continuum subtractions:
\begin{eqnarray}
&&\Pi _{1}(\mathbf{M}^{2},\mathbf{s}_{0},q^{2})=%
\int_{(m_{b}+m_{c})^{2}}^{s_{0}}ds\int_{m_{b}^{2}}^{s_{0}^{\prime
}}ds^{\prime }\rho _{1}(s,s^{\prime },q^{2})  \notag \\
&&\times e^{-s/M_{1}^{2}-s^{\prime }/M_{2}^{2}}.  \label{eq:CorrF1}
\end{eqnarray}%
The spectral density $\rho _{1}(s,s^{\prime },q^{2})$ is evaluated
as the imaginary part of the amplitude $\Pi _{1}^{\mathrm{OPE}%
}(p^{2},p^{\prime 2},q^{2})$. Computations are performed by taking unto
account nonperturbative terms up to dimension $10$. In general, $\rho
_{1}(s,s^{\prime },q^{2})$ depends on the parameters $\mathbf{M}%
^{2}=(M_{1}^{2},M_{2}^{2})$, $\mathbf{s}_{0}=(s_{0},s_{0}^{\prime })$ and $%
q^{2}$. The pairs $(M_{1}^{2},s_{0})$ and $(M_{2}^{2},s_{0}^{\prime })$
correspond to the hybrid $H_{\mathrm{bc}}$ and $B^{\ast }{}^{0}$ channels,
respectively.

In the case under analysis $\rho _{1}(s,s^{\prime },q^{2})$ has the form
\begin{eqnarray}
\rho _{1}(s,s^{\prime },q^{2}) &=&\rho _{1}^{\mathrm{pert.}}(s^{\prime
})+\rho _{1\mathrm{a}}^{\mathrm{Dim4}}(s^{\prime })+\rho _{1\mathrm{b}}^{%
\mathrm{Dim4}}(s^{\prime },q^{2}),  \notag \\
&&
\end{eqnarray}%
i.e., only the perturbative and dimension-$4$ terms contribute to the
correlation function $\Pi _{1}(\mathbf{M}^{2},\mathbf{s}_{0},q^{2})$. It is
worth to emphasize that a limited number of the terms in $\rho
_{1}(s,s^{\prime },q^{2})$ in this process is connected with the forms of
the interpolating currents for $H_{\mathrm{bc}}$, $B^{\ast 0}$, and $D^{+}$,
mathematical operations applied to derive $\rho _{1}(s,s^{\prime },q^{2})$,
as well as with the approximation $m_{u}=m_{d}=0$ adopted in the present
work.

The components of $\rho _{1}(s,s^{\prime },q^{2})$ are given by the
expressions:%
\begin{eqnarray}
&&\rho _{1}^{\mathrm{pert.}}(s^{\prime })=\frac{g_{s}^{2}m_{b}m_{c}}{32\pi
^{2}}\int_{0}^{1}d\alpha \int_{0}^{1-\alpha }d\beta \frac{\theta (N)N^2}{%
\alpha ^{2}\beta (\alpha +\beta -1)},  \notag \\
&&\rho _{1\mathrm{a}}^{\mathrm{Dim4}}(s^{\prime })=\frac{g_{s}^{2}m_{b}m_{c}%
}{144}\langle \alpha _{s}G^{2}/\pi \rangle \int_{0}^{1}d\alpha
\int_{0}^{1-\alpha }d\beta \theta (N)  \notag \\
&&\times \frac{(\alpha ^{2}+\beta ^{2})}{\alpha ^{2}\beta ^{2}(\alpha +\beta
-1)},  \notag \\
&&\rho _{1\mathrm{b}}^{\mathrm{Dim4}}(s^{\prime },q^{2})=\frac{m_{b}m_{c}\pi
^{2}}{4}\langle \alpha _{s}G^{2}/\pi \rangle \int_{0}^{1}d\alpha
\int_{0}^{1-\alpha }d\beta \theta (\widetilde{N})  \notag \\
&&\times \frac{1}{\alpha \beta ^{2}}\left[ \frac{(\alpha +\beta -1)}{3\alpha
}+\frac{1}{2}\right] .  \label{eq:Rho}
\end{eqnarray}%
Here
\begin{eqnarray}
N &=&\frac{s^{\prime }(1-\alpha -\beta )-m_{b}^{2}\alpha -m_{c}^{2}\beta }{%
\beta },  \notag \\
\widetilde{N} &=&\frac{q^{2}(1-\alpha -\beta )-m_{b}^{2}\alpha
-m_{c}^{2}\beta }{\beta },
\end{eqnarray}%
where $\alpha $ and $\beta $ are the Feynman parameters.

In numerical computations, we choose the parameters $(M_{1}^{2},s_{0})$ and $%
(M_{2}^{2},s_{0}^{\prime })$ in accordance with the scheme: In the hybrid
meson channel we use $(M_{1}^{2},s_{0})$ from Eq.\ (\ref{eq:Wind1}), whereas
for the channel of the $B^{\ast }{}^{0}$ meson employ
\begin{equation}
M_{2}^{2}\in \lbrack 5.5,6.5]~\mathrm{GeV}^{2},\ s_{0}^{\prime }\in \lbrack
34,35]~\mathrm{GeV}^{2}.
\end{equation}

\begin{figure}[h]
\includegraphics[width=8.5cm]{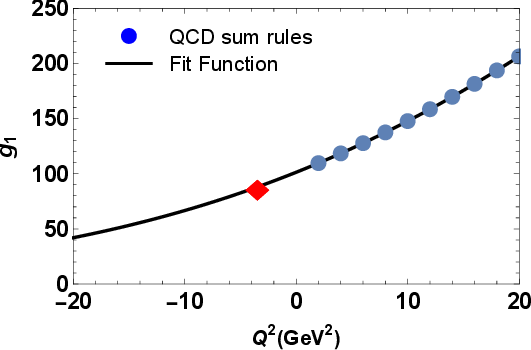}
\caption{QCD data and extrapolating function $\mathcal{G}_{1}(Q^{2})$ . The
red diamond fixes the point $Q^{2}=-m_{D}^{2}$. }
\label{fig:Fit}
\end{figure}

The SR method allows one to compute the form factor in the $q^{2}<0$ domain.
But we know that the coupling $g_{1}$ should be determined at the mass shell
of the $D^{+}$ meson, i.e., $g_{1}=g_{1}(m_{D}^{2})$. To escape from these
problems, it is convenient to use a variable $Q^{2}=-q^{2}$ and denote a new
function as $g_{1}(Q^{2})$. Then we calculate the form factor $g_{1}(Q^{2})$
at $Q^{2}=2-20~\mathrm{GeV}^{2}$. The results of these computations are
plotted in Fig.\ \ref{fig:Fit}. Later we use the fit function $\mathcal{G}%
_{1}(Q^{2})$ which at momenta $Q^{2}>0$ generates the same QCD data, but can
be extrapolated to a region of negative $Q^{2}$. It is convenient to employ
the function
\begin{equation}
\mathcal{G}_{1}(Q^{2})=\mathcal{G}_{1}^{0}\mathrm{\exp }\left[ c_{1}^{1}%
\frac{Q^{2}}{m^{2}}+c_{1}^{2}\left( \frac{Q^{2}}{m^{2}}\right) ^{2}\right] ,
\label{eq:FitF}
\end{equation}%
where $\mathcal{G}_{1}^{0}$, $c_{1}^{1}$, and $c_{1}^{2}$ are fitted
constants. Then, from comparison of the SR data and $\mathcal{G}_{1}(Q^{2})$%
, we find
\begin{equation}
\mathcal{G}_{1}^{0}=101.37,\text{ }c_{1}^{1}=2.07,\text{and }c_{1}^{2}=-0.58.
\label{eq:FF1}
\end{equation}%
The function $\mathcal{G}_{1}(Q^{2},m^{2})$ is demonstrated in Fig.\ \ref%
{fig:Fit} as well, one sees nice agreement with the SR data. Then for the $%
g_{1}$, we get
\begin{equation}
g_{1}\equiv \mathcal{G}_{1}(-m_{D}^{2})=87.9\pm 16.5.  \label{eq:g1}
\end{equation}

The width of the decay $H_{\mathrm{bc}}\rightarrow B^{\ast +}D^{0}$ is found
by means of the formula%
\begin{equation}
\Gamma \lbrack H_{\mathrm{bc}}\rightarrow B^{\ast +}D^{0}]=g_{1}^{2}\frac{%
\lambda _{1}}{40\pi ^{2}m^{2}}|M|^{2},  \label{eq:PDw2}
\end{equation}%
where $|M|^{2}$ is%
\begin{eqnarray}
&&|M|^{2}=\frac{1}{24m^{4}m_{B^{\ast }}^{2}}\left[
m^{8}-2m^{2}(2m_{D}^{2}-3m_{B^{\ast }}^{2})\right.  \notag \\
&&\times (m_{B^{\ast }}^{2}-m_{D}^{2})^{2}+(m_{B^{\ast
}}^{2}-m_{D}^{2})^{4}+m^{6}(6m_{B^{\ast }}^{2}-4m_{D}^{2})  \notag \\
&&\left. +2m^{4}(3m_{D}^{4}-8m_{B^{\ast }}^{2}m_{D}^{2}-7m_{B^{\ast }}^{2})
\right] .
\end{eqnarray}%
In Eq.\ (\ref{eq:PDw2}) $\lambda _{1}$ is equal to $\lambda (m,m_{B^{\ast
}},m_{D})$, where
\begin{eqnarray}
\lambda (a,b,c) &=&\frac{\sqrt{%
a^{4}+b^{4}+c^{4}-2(a^{2}b^{2}+a^{2}c^{2}+b^{2}c^{2})}}{2a}.  \notag \\
&&
\end{eqnarray}

Computations yield for the partial width of this process $\ $%
\begin{equation}
\Gamma \left[ H_{\mathrm{bc}}\rightarrow B^{\ast +}D^{0}\right] =(25.4\pm
6.8\mp 1.2)~\mathrm{MeV}.  \label{eq:DW2}
\end{equation}%
The first uncertainties above are generated by the coupling $g_{1}$, while
the second ones are connected with errors in the mesons' masses in Eq.\ (\ref%
{eq:PDw2}).

Then the full width of the hybrid meson can be estimated within limits

\begin{equation}
\Gamma \left[ H_{\mathrm{bc}}\right] =(50.8\pm 9.8)~\mathrm{MeV}.
\end{equation}


\section{Full width of $\widetilde{H}_{\mathrm{bc}}$}

\label{sec:TildeHbcWidth}

The hybrid meson $\widetilde{H}_{\mathrm{bc}}$ with the mass $\widetilde{m}%
=(7.685\pm 0.040)~\mathrm{GeV}$ and $J^{\mathrm{P}}=2^{+}$ has numerous
decay channels. These parameters enable one to fix its two-meson decay
modes. One can be easily convinced that processes $\widetilde{H}_{\mathrm{bc}%
}\rightarrow B^{+}D^{0}$, $B^{0}D^{+}$, $B^{\ast +}D^{\ast 0}$, $B^{\ast
0}D^{\ast +}$, $B_{s}^{0}D_{s}^{+}$, and $B_{s}^{\ast 0}D_{s}^{\ast +}$ are
kinematically allowed decay modes of $\widetilde{H}_{\mathrm{bc}}$. In our
computations, we take into account that in approximation adopted in the
present work decays to $B^{+}D^{0}$ and $B^{0}D^{+}$, $B^{\ast +}D^{\ast 0}$
and $B^{\ast 0}D^{\ast +}$ have the same widths.


\subsection{Decays $\widetilde{H}_{\mathrm{bc}}\rightarrow B^{+}D^{0}$ and $%
\widetilde{H}_{\mathrm{bc}}\rightarrow B^{0}D^{+}$}


To find the width of the process $\widetilde{H}_{\mathrm{bc}}\rightarrow
B^{+}D^{0}$, we begin from analysis of the correlation function
\begin{eqnarray}
\widetilde{\Pi }_{\alpha \beta }(p,p^{\prime }) &=&i^{2}\int
d^{4}xd^{4}ye^{ip^{\prime }x}e^{iqy}\langle 0|\mathcal{T}\{J^{B^{+}}(x)
\notag \\
&&\times J^{D^{0}}(y)\widetilde{J}_{\alpha \beta }^{\dagger }(0)\}|0\rangle ,
\label{eq:CF4}
\end{eqnarray}%
in order to find the sum rule for the form factor $\widetilde{g}_{1}(q^{2})$
and estimate the coupling $\widetilde{g}_{1}$ describing the strong
interaction of particles at the vertex $\widetilde{H}_{\mathrm{bc}%
}B^{+}D^{0} $. In Eq.\ (\ref{eq:CF4}), $J^{B^{+}}(x)$ and $J^{D^{0}}(y)$ are
the interpolating currents of the pseudoscalar mesons $B^{+}$ and $D^{0}$
\begin{equation}
J^{B^{+}}(x)=\overline{b}_{i}(x)i\gamma _{5}u_{i}(x),\ J^{D^{0}}(x)=%
\overline{u}_{j}(x)i\gamma _{5}c_{j}(x).
\end{equation}

The physical side of this SR is given by the formula
\begin{eqnarray}
&&\widetilde{\Pi }_{\alpha \beta }^{\mathrm{Phys}}(p,p^{\prime })=\frac{%
\langle 0|J^{B^{+}}|B^{+}(p^{\prime })\rangle }{p^{\prime 2}-m_{B}^{2}}\frac{%
\langle 0|J^{D^{0}}|D^{0}(q)\rangle }{q^{2}-m_{D^{0}}^{2}}  \notag \\
&&\times \langle B^{+}(p^{\prime })D^{0}(q)|\widetilde{H}_{\mathrm{bc}%
}(p,\epsilon )\rangle \frac{\langle \widetilde{H}_{\mathrm{bc}}(p,\epsilon )|%
\widetilde{J}_{\alpha \beta }^{\dagger }|0\rangle }{p^{2}-\widetilde{m}^{2}}
\notag \\
&&+\cdots ,
\end{eqnarray}%
where $m_{D^{0}}$ and $m_{B^{+}}$ are the masses of the mesons $D^{0}$ and $%
B^{+}$, respectively:
\begin{eqnarray}
m_{B^{+}} &=&(5279.42\pm 0.08)~\mathrm{MeV},  \notag \\
m_{D^{0}} &=&(1864.84\pm 0.05)~\mathrm{MeV}.  \label{eq:ME2}
\end{eqnarray}

The matrix elements of the $D^{0}$ and $B^{+}$ mesons have the forms $%
f_{D}m_{D^{0}}^{2}/m_{c}$ and $f_{B}m_{B^{+}}^{2}/m_{b}$. \ The decay
constant $f_{D}$ of the $D^{0}$ meson has been introduced above, whereas for
the decay constant of the meson $B^{+}$ we use $f_{B}=(206\pm 7)\ \mathrm{%
MeV.}$ The vertex $\langle B^{+}(p^{\prime })D^{0}(q)|\widetilde{H}_{\mathrm{%
bc}}(p,\epsilon )\rangle $ is introduced by the formula
\begin{equation}
\langle B^{+}(p^{\prime })D^{0}(q)|\widetilde{H}_{\mathrm{bc}}(p,\epsilon
)\rangle =\widetilde{g}_{1}(q^{2})\epsilon _{\mu \nu }(p)p^{\prime \mu
}q^{\nu }.
\end{equation}%
Then, the correlator becomes equal to
\begin{eqnarray}
&&\widetilde{\Pi }_{\alpha \beta }^{\mathrm{Phys}}(p,p^{\prime })=\frac{%
\widetilde{g}_{1}(q^{2})\widetilde{\Lambda }f_{D}m_{D}^{2}f_{B}m_{B^{+}}^{2}%
}{m_{b}m_{c}\left( p^{2}-\widetilde{m}^{2}\right) (p^{\prime
2}-m_{B^{+}}^{2})}  \notag \\
&&\times \frac{1}{(q^{2}-m_{D^{0}}^{2})}\left\{ \frac{\widetilde{\lambda }%
^{2}}{3}g_{\alpha \beta }+\left[ \frac{m_{B^{+}}^{2}}{\widetilde{m}^{2}}+%
\frac{2\widetilde{\lambda }^{2}}{3\widetilde{m}^{2}}\right] p_{\alpha
}p_{\beta }\right.  \notag \\
&&\left. +p_{\alpha }^{\prime }p_{\beta }^{\prime }-\frac{\widetilde{m}%
^{2}+m_{B^{+}}^{2}-q^{2}}{2\widetilde{m}^{2}}(p_{\alpha }p_{\beta }^{\prime
}+p_{\beta }p_{\alpha }^{\prime })\right\} ,
\end{eqnarray}%
where $\widetilde{\lambda }=\lambda (\widetilde{m},m_{B^{+}},q)$.

In terms of the quark-gluon propagators the correlation function $\widetilde{%
\Pi }_{\alpha \beta }(p,p^{\prime })$ reads%
\begin{eqnarray}
&&\Pi _{\alpha \beta }^{\mathrm{OPE}}(p,p^{\prime })=\int
d^{4}xd^{4}ye^{ip^{\prime }x}e^{iqy}g_{s}\frac{{\lambda }_{ab}^{n}}{2}%
\widetilde{G}_{\rho \alpha }^{n}(0)  \notag \\
&&\times \mathrm{Tr}\left[ \gamma _{5}S_{u}^{ij}(x-y)\gamma
_{5}S_{c}^{jb}(y)\gamma _{5}\sigma _{\beta }^{\rho }S_{b}^{ai}(-x)\right] .
\end{eqnarray}%
The SR for the form factor $\widetilde{g}_{1}(q^{2})$ is derived by
employing invariant amplitudes $\widetilde{\Pi }_{1}^{\mathrm{Phys}%
}(p^{2},p^{\prime 2},q^{2})$ and $\widetilde{\Pi }_{1}^{\mathrm{OPE}%
}(p^{2},p^{\prime 2},q^{2})$ which correspond to terms $g_{\alpha \beta }$
in the correlation functions
\begin{eqnarray}
&&\widetilde{g}_{1}(q^{2})=\frac{3m_{b}m_{c}(q^{2}-m_{D^{0}}^{2})}{%
\widetilde{\Lambda }f_{D}m_{D^{0}}^{2}f_{B}m_{B^{+}}^{2}\widetilde{\lambda }%
^{2}}e^{m^{2}/M_{1}^{2}}e^{m_{B^{+}}^{2}/M_{2}^{2}}  \notag \\
&&\times \widetilde{\Pi }_{1}(\mathbf{M}^{2},\mathbf{s}_{0},q^{2}).
\end{eqnarray}%
The function $\widetilde{\Pi }_{1}(\mathbf{M}^{2},\mathbf{s}_{0},q^{2})$ is
given by the same expression Eq.\ (\ref{eq:CorrF1}) but with new spectral
density $\rho _{2}(s,s^{\prime },q^{2})$. In this decay $\rho
_{2}(s,s^{\prime },q^{2})$ has perturbative, dimension-3,-5 and-7
components. We do not write down their explicit expressions, but note that
nonperturbative terms are proportional to light quark condensates $\langle
\overline{q}q\rangle $, $\langle \overline{q}g_{s}\sigma Gq\rangle $, and $%
\langle \alpha _{s}G^{2}/\pi \rangle \langle \overline{q}q\rangle $. \ In
numerical computations effects of these terms are also taken into account.
Below we present their values:

\begin{eqnarray}
&&\langle \overline{q}q\rangle =-(0.24\pm 0.01)^{3}~\mathrm{GeV}^{3},\
\langle \overline{s}s\rangle =(0.8\pm 0.1)\langle \overline{q}q\rangle ,
\notag \\
&&\langle \overline{q}g_{s}\sigma Gq\rangle =m_{0}^{2}\langle \overline{q}%
q\rangle ,\ m_{0}^{2}=(0.8\pm 0.1)~\mathrm{GeV}^{2},  \notag \\
&&\langle \overline{s}g_{s}\sigma Gs\rangle =m_{0}^{2}\langle \overline{s}%
s\rangle .  \label{eq:Parameters}
\end{eqnarray}%
 Eq.\ (\ref{eq:Parameters}) contains also vacuum expectation values of $s$
quark operators that appear while considering decays of \ $\widetilde{H}_{%
\mathrm{bc}}$ to strange mesons.

For the Borel and continuum subtraction parameters $(M_{1}^{2},s_{0})$ in
the hybrid $\widetilde{H}_{\mathrm{bc}}$ meson's channel we use ones from
Eq.\ (\ref{eq:Wind1A}). In the $B^{+}$ meson's channel, we employ%
\begin{equation}
M_{2}^{2}\in \lbrack 5.5,6.5]~\mathrm{GeV}^{2},\ s_{0}^{\prime }\in \lbrack
33.5,34.5]~\mathrm{GeV}^{2}.
\end{equation}

\begin{figure}[h]
\includegraphics[width=8.5cm]{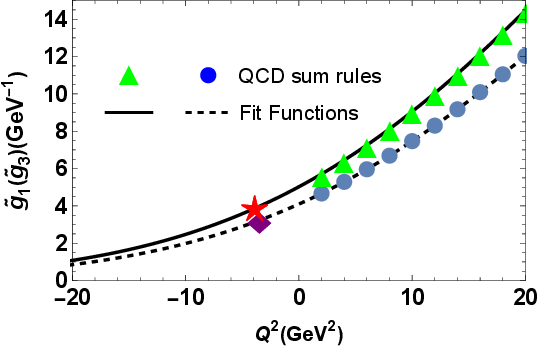}
\caption{SR data and functions $\widetilde{\mathcal{G}}_{1}(Q^{2})$ (dashed
line) and $\widetilde{\mathcal{G}}_{3}(Q^{2})$ (solid line). The labels are
fixed at the points $Q^{2}=-m_{D^0}^{2}$ and $Q^{2}=-m_{D_s}^{2}$. }
\label{fig:Fit1}
\end{figure}

The coupling $\widetilde{g}_{1}$ is extracted at the $D^{0}$ meson mass
shell $q^{2}=m_{D^{0}}^{2}$ by means of the extrapolating function $%
\widetilde{\mathcal{G}}_{1}(Q^{2})$ with parameters $\widetilde{\mathcal{G}}%
_{1}^{0}=4.10\ \mathrm{GeV}^{-1}$, $\widetilde{c}_{1}^{1}=3.91,$and $%
\widetilde{c}_{1}^{2}=-2.16$ (see, Fig.\ \ref{fig:Fit1}). Note that the
functions $\widetilde{\mathcal{G}}_{i}(Q^{2})$ are determined after the
replacement $m\rightarrow \widetilde{m}$ by  Eq.\ (\ref{eq:FitF}).

As a result, we get
\begin{equation}
\widetilde{g}_{1}\equiv \widetilde{\mathcal{G}}_{1}(-m_{D^{0}}^{2})=(3.23\pm
0.61)\ \mathrm{GeV}^{-1}.
\end{equation}%
The partial width of the decay $\widetilde{H}_{\mathrm{bc}}\rightarrow
B^{+}D^{0}$ is equal to
\begin{equation}
\Gamma \left[ \widetilde{H}_{\mathrm{bc}}\rightarrow B^{+}D^{0}\right] =%
\widetilde{g}_{1}^{2}\frac{\widetilde{\lambda }_{1}}{40\pi \widetilde{m}^{2}}%
|\widetilde{M}_{1}|^{2},  \label{eq:PW1}
\end{equation}%
where $\widetilde{\lambda }_{1}=\lambda (\widetilde{m},m_{B^{+}},m_{D^{0}})$
and
\begin{eqnarray}
&&|\widetilde{M}_{1}|^{2}=\frac{1}{24\widetilde{m}^{4}}\left\{ \widetilde{m}%
^{8}+(m_{B^{+}}^{2}-m_{D^{0}}^{2})^{4}\right.  \notag \\
&&\left. +4\widetilde{m}%
^{4}m_{D^{0}}^{2}(m_{B^{+}}^{2}+3m_{D^{0}}^{2})-(m_{B^{+}}^{2}+7m_{D^{0}}^{2})\right.
\notag \\
&&\left. \times \lbrack \widetilde{m}^{6}+\widetilde{m}%
^{2}(m_{B^{+}}^{2}-m_{D^{0}}^{2})^{2}]\right\} .
\end{eqnarray}%
Then it is easy to evaluate
\begin{equation}
\Gamma \left[ \widetilde{H}_{\mathrm{bc}}\rightarrow B^{+}D^{0}\right]
=(38.0\pm 10.2\pm 4.2)~\mathrm{MeV},
\end{equation}%
where the first errors are connected with the uncertainties in the coupling $%
\widetilde{g}_{1}$, whereas the second one are generated by the ambiguities
in the masses $\widetilde{m}$ and $m_{B^{+}}$, and $m_{D^{0}}$.

The width of the process $\widetilde{H}_{\mathrm{bc}}\rightarrow B^{0}D^{+}$
is equal to Eq.\ (\ref{eq:PW1}) provided one neglects the mass differences
between charged and neutral $B$ and $D$ mesons.

\subsection{ $\widetilde{H}_{\mathrm{bc}}\rightarrow B^{\ast +}D^{\ast 0}$,
and $B^{\ast 0}D^{\ast +}$}


In this subsection, we analyze the decay $\widetilde{H}_{\mathrm{bc}%
}\rightarrow B^{\ast +}D^{\ast 0}$ bearing in mind that the second process
within our approximation is characterized by the same width. To explore the
process $\widetilde{H}_{\mathrm{bc}}\rightarrow B^{\ast +}D^{\ast 0}$, we
consider the correlator
\begin{eqnarray}
\widetilde{\Pi }_{\mu \nu \alpha \beta }(p,p^{\prime }) &=&i^{2}\int
d^{4}xd^{4}ye^{ip^{\prime }y}e^{iqx}\langle 0|\mathcal{T}\{J_{\mu }^{B^{\ast
+}}(x)  \notag \\
&&\times J_{\nu }^{D^{\ast 0}}(y)\widetilde{J}_{\alpha \beta }^{\dagger
}(0)\}|0\rangle ,  \label{eq:CF7}
\end{eqnarray}%
that enables us to derive the SR for the strong form factor $\widetilde{g}%
_{2}(q^{2})$ at the vertex $\widetilde{H}_{\mathrm{bc}}B^{\ast +}D^{\ast 0}$%
.  Here, $J_{\mu }^{B^{\ast +}}(x)$ and $J_{\nu }^{D^{\ast 0}}(x)$ are the $%
B^{\ast +}$ and $D^{\ast }{}^{0}$ mesons' interpolating currents. They are
defined by the following manner
\begin{equation}
J_{\mu }^{B^{\ast +}}(x)=\overline{b}_{i}(x)\gamma _{\mu }u_{i}(x),\ J_{\nu
}^{D^{\ast 0}}(x)=\overline{u}_{j}(x)\gamma _{\nu }c_{j}(x).
\end{equation}

The correlator $\widetilde{\Pi }_{\mu \nu \alpha \beta }(p,p^{\prime })$
obtained using the physical parameters of the particles $\widetilde{H}_{%
\mathrm{bc}}$, $B^{\ast +}$, and $D^{\ast 0}$ is
\begin{eqnarray}
&&\widetilde{\Pi }_{\mu \nu \alpha \beta }^{\mathrm{Phys}}(p,p^{\prime })=%
\frac{\langle 0|J_{\mu }^{B^{\ast +}}|B^{\ast +}(p^{\prime },\varepsilon
_{1})\rangle }{p^{\prime 2}-m_{B^{\ast +}}^{2}}\frac{\langle 0|J_{\nu
}^{D^{\ast 0}}|D^{\ast 0}(q,\varepsilon _{2})\rangle }{q^{2}-m_{D^{\ast
0}}^{2}}  \notag \\
&&\times \langle B^{\ast +}(p^{\prime },\varepsilon _{1})D^{\ast
0}(q,\varepsilon _{2})|\widetilde{H}_{\mathrm{bc}}(p,\epsilon )\rangle \frac{%
\langle \widetilde{H}_{\mathrm{bc}}(p,\epsilon )|\widetilde{J}_{\alpha \beta
}^{\dagger }|0\rangle }{p^{2}-\widetilde{m}^{2}}  \notag \\
&&+\cdots ,  \label{eq:PhysSide1}
\end{eqnarray}%
where $\varepsilon _{1\mu }$ and $\varepsilon _{2\mu }$ are the polarization
vectors of the $B^{\ast +}$ and $D^{\ast }{}^{0}$ mesons, respectively. In
Eq.\ (\ref{eq:PhysSide1}) $m_{B^{\ast +}}$ and $m_{D^{\ast 0}}$ are the
masses of these mesons%
\begin{eqnarray}
m_{B^{\ast +}} &=&(5324.75\pm 0.2)~\mathrm{MeV},  \notag \\
m_{D^{\ast 0}} &=&(2006.85\pm 0.05)~\mathrm{MeV}.
\end{eqnarray}%
The matrix elements which we are used are
\begin{eqnarray}
\langle 0|J_{\mu }^{B^{\ast +}}|B^{\ast +}(p^{\prime },\varepsilon
_{1})\rangle  &=&f_{B^{\ast }}m_{B^{\ast +}}\varepsilon _{1\mu }(p^{\prime
}),  \notag \\
\langle 0|J_{\nu }^{D^{\ast 0}}|D^{\ast 0}(q,\varepsilon _{2})\rangle
&=&f_{D^{\ast }}m_{D^{\ast 0}}\varepsilon _{2\nu }(p^{\prime }),
\end{eqnarray}%
where  $f_{D^{\ast }}=(252.2\pm 22.66)\ \mathrm{MeV}$ is the decay constant
of the meson  $D^{\ast }{}^{0}$.

The vertex $\langle B^{\ast +}(p^{\prime },\varepsilon _{1})D^{\ast
0}(q,\varepsilon _{2})|\widetilde{H}_{\mathrm{c}}(p,\epsilon )\rangle $ is
\begin{eqnarray}
&&\langle B^{\ast +}(p^{\prime },\varepsilon _{1})D^{\ast 0}(q,\varepsilon
_{2})|\widetilde{H}_{\mathrm{bc}}(p,\epsilon )\rangle =\widetilde{g}%
_{2}(q^{2})\epsilon _{\tau \rho }\left[ \varepsilon _{1}^{\ast }\cdot
q\right.  \notag \\
&&\left. \times \varepsilon _{2}^{\tau \ast }p^{\prime \rho }+\varepsilon
_{2}^{\ast }\cdot p^{\prime }\varepsilon _{1}^{\ast \tau }q^{\rho
}-p^{\prime }\cdot q\varepsilon _{1}^{\tau \ast }\varepsilon _{2}^{\rho \ast
}-\varepsilon _{1}^{\ast }\cdot \varepsilon _{2}^{\ast }p^{\prime \tau
}q^{\rho }\right] .  \notag \\
&&  \label{eq:TVVVertex}
\end{eqnarray}%
Then, we get for $\widetilde{\Pi }_{\mu \nu \alpha \beta }^{\mathrm{Phys}%
}(p,p^{\prime })$
\begin{eqnarray}
&&\widetilde{\Pi }_{\mu \nu \alpha \beta }^{\mathrm{Phys}}(p,p^{\prime })=%
\frac{\widetilde{g}_{2}(q^{2})\widetilde{\Lambda }f_{B^{\ast }}f_{D^{\ast
}}m_{B^{\ast +}}m_{D^{\ast 0}}}{\left( p^{2}-\widetilde{m}^{2}\right)
(p^{\prime 2}-m_{B^{\ast +}}^{2})(q^{2}-m_{D^{\ast 0}}^{2})}  \notag \\
&&\times \left\{ \frac{m_{B^{\ast +}}^{4}-2m_{B^{\ast +}}^{2}(2\widetilde{m}%
^{2}+q^{2})+(\widetilde{m}^{2}-q^{2})(3\widetilde{m}^{2}-q^{2})}{12%
\widetilde{m}^{2}}\right.  \notag \\
&&\times g_{\mu \nu }g_{\alpha \beta }+\frac{1}{3}g_{\alpha \beta }\left(
\frac{m_{B^{\ast +}}^{2}}{\widetilde{m}^{2}}p_{\mu }p_{\nu }+2p_{\mu
}^{\prime }p_{\nu }^{\prime }\right) -\frac{1}{6\widetilde{m}^{2}}g_{\alpha
\beta }  \notag \\
&&\times \left[ (m_{B^{\ast +}}^{2}+3\widetilde{m}^{2}-q^{2})p_{\mu }p_{\nu
}^{\prime }+(m_{B^{\ast +}}^{2}+\widetilde{m}^{2}-q^{2})p_{\mu }^{\prime
}p_{\nu }\right]  \notag \\
&&\left. +\text{other structures}\right\} .  \label{eq:PhysSide2}
\end{eqnarray}

Calculations for $\widetilde{\Pi }_{\mu \nu \alpha \beta }^{\mathrm{OPE}%
}(p,p^{\prime })$ yield
\begin{eqnarray}
&&\widetilde{\Pi }_{\mu \nu \alpha \beta }^{\mathrm{OPE}}(p,p^{\prime
})=i^{2}\int d^{4}xd^{4}ye^{ip^{\prime }x}e^{iqy}g_{s}\frac{{\lambda }%
_{ab}^{n}}{2}\widetilde{G}_{\rho \alpha }^{n}(0)  \notag \\
&&\times \mathrm{Tr}\left[ \gamma _{\mu }S_{u}^{ij}(x-y)\gamma _{\nu
}S_{c}^{jb}(y)\gamma _{5}\sigma _{\beta }^{\rho }S_{b}^{ai}(-x)\right].
\label{eq:CF8}
\end{eqnarray}%
The SR for the form factor $\widetilde{g}_{2}(q^{2})$ is derived using the
structures $g_{\mu \nu }g_{\alpha \beta }$ in the correlation functions. The
amplitude $\widetilde{\Pi }_{2}^{\mathrm{OPE}}(p^{2},p^{\prime 2},q^{2})$
that corresponds to this structure receives contributions from the
perturbative, dimension-5 and -7 terms.

In numerical analysis, the parameters $M_{2}^{2}$ and $s_{0}^{\prime }$ in
the $B^{\ast +}$ meson channel are chosen in the form
\begin{equation}
M_{2}^{2}\in \lbrack 5.5,6.5]~\mathrm{GeV}^{2},\ s_{0}^{\prime }\in \lbrack
34,35]~\mathrm{GeV}^{2}.
\end{equation}%
The strong coupling $\widetilde{g}_{2}$ amounts to
\begin{equation}
\widetilde{g}_{2}=\widetilde{\mathcal{G}}_{2}(-m_{D^{\ast 0}}^{2})=(0.51\pm
0.11)\ \mathrm{GeV}^{-1}.
\end{equation}%
It has been estimated at the mass shell $q^{2}=m_{D^{\ast 0}}^{2}$ of the $%
D^{\ast 0}$ meson by employing the extrapolating function $\widetilde{%
\mathcal{G}}_{2}(Q^{2})$ with the parameters $\widetilde{\mathcal{G}}%
_{2}^{0}=0.54\ \mathrm{GeV}^{-1},$ $\widetilde{c}_{2}^{1}=0.74,$and $%
\widetilde{c}_{2}^{2}=-0.48$.

The width of this decay is
\begin{equation}
\Gamma \left[ \widetilde{H}_{\mathrm{bc}}\rightarrow B^{\ast +}D^{\ast 0}%
\right] =\widetilde{g}_{2}^{2}\frac{\widetilde{\lambda }_{2}}{40\pi
\widetilde{m}^{2}}|\widetilde{M}_{2}|^{2},
\end{equation}%
with $\widetilde{\lambda }_{2}$ being equal to $\lambda (\widetilde{m}%
,m_{B^{\ast +}},m_{D^{\ast 0}})$. Here%
\begin{eqnarray}
&&|\widetilde{M}_{2}|^{2}=\frac{1}{12\widetilde{m}^{4}}\left[ 6\widetilde{m}%
^{8}+9\widetilde{m}^{6}(m_{B^{\ast +}}^{2}+m_{D^{\ast 0}}^{2})\right.  \notag
\\
&&+\widetilde{m}^{4}\left( m_{B^{\ast +}}^{4}+34m_{B^{\ast +}}^{2}m_{D^{\ast
0}}^{2}+m_{D^{\ast 0}}^{4}\right) +\widetilde{m}^{2}\left( m_{B^{\ast
+}}^{4}\right.  \notag \\
&&\left. \left. -m_{D^{\ast 0}}^{4}\right) (m_{B^{\ast +}}^{2}-m_{D^{\ast
0}}^{2})+(m_{B^{\ast +}}^{2}-m_{D^{\ast 0}}^{2})^{4}\right] . \\
&&
\end{eqnarray}%
Then, we get
\begin{equation}
\Gamma \left[ \widetilde{H}_{\mathrm{bc}}\rightarrow B^{\ast +}D^{\ast 0}%
\right] =(26.8\pm 8.2\pm 2.0)~\mathrm{MeV}.
\end{equation}


\subsection{Processes $\widetilde{H}_{\mathrm{bc}}\rightarrow
B_{s}^{0}D_{s}^{+}$, and $B_{s}^{\ast 0}D_{s}^{\ast +}$}


The tensor hybrid structure $\widetilde{H}_{\mathrm{bc}}$ can also decay to
strange mesons $B_{s}^{0}D_{s}^{+}$ and \ $B_{s}^{\ast 0}D_{s}^{\ast +}$.
These modes differ from those considered above by some matrix elements and
parameters of the final-state mesons $B_{s}^{0(\ast )}$ and $D_{s}^{+(\ast
)} $.

In the case of the decay $\widetilde{H}_{\mathrm{bc}}\rightarrow
B_{s}^{0}D_{s}^{+}$, the correlation function that should be analyzed is
\begin{eqnarray}
\widehat{\Pi }_{\alpha \beta }(p,p^{\prime }) &=&i^{2}\int
d^{4}xd^{4}ye^{ip^{\prime }x}e^{iqy}\langle 0|\mathcal{T}\{J^{B_{s}}(x)
\notag \\
&&\times J^{D_{s}^{+}}(y)\widetilde{J}_{\alpha \beta }^{\dagger
}(0)\}|0\rangle ,
\end{eqnarray}%
where $J^{B_{s}^{0}}(x)$ and $J^{D_{s}^{+}}(x)$ are the interpolating
currents of the mesons $B_{s}^{0}$ and $D_{s}^{+}$
\begin{equation}
J^{B_{s}^{0}}(x)=\overline{b}_{i}(x)i\gamma _{5}s_{i}(x),\ J^{D_{s}^{+}}(x)=%
\overline{s}_{j}(x)i\gamma _{5}c_{j}(x).
\end{equation}

To write $\widehat{\Pi }_{\alpha \beta }(p,p^{\prime })$ in terms of the
physical parameters of the particles $\widetilde{H}_{\mathrm{bc}}$, $%
B_{s}^{0}$, and $D_{s}^{+}$, we recast it into the form
\begin{eqnarray}
&&\widehat{\Pi }_{\alpha \beta }^{\mathrm{Phys}}(p,p^{\prime })=\frac{%
\langle 0|J^{B_{s}^{0}}|B_{s}^{0}(p^{\prime })\rangle }{p^{\prime
2}-m_{B_{s}}^{2}}\frac{\langle 0|J^{D_{s}^{+}}|D_{s}^{+}(q)\rangle }{%
q^{2}-m_{D_{s}}^{2}}  \notag \\
&&\times \langle B_{s}^{0}(p^{\prime })D_{s}^{+}(q)|\widetilde{H}_{\mathrm{bc%
}}(p,\epsilon )\rangle \frac{\langle \widetilde{H}_{\mathrm{bc}}(p,\epsilon
)|\widetilde{J}_{\alpha \beta }^{\dagger }|0\rangle }{p^{2}-\widetilde{m}^{2}%
}  \notag \\
&&+\cdots ,
\end{eqnarray}%
with $m_{B_{s}}=(5366.93\pm 0.10)~\mathrm{MeV}$ and $m_{D_{s}}=(1968.35\pm
0.07)~\mathrm{MeV}$ being the masses of the mesons $B_{s}^{0}$ and $%
D_{s}^{+} $, respectively. The matrix elements of these pseudoscalar mesons
required for further analysis are given by the formulas
\begin{eqnarray}
\langle 0|J^{D_{s}^{+}}|D_{s}^{+}(q)\rangle &=&\frac{f_{D_{s}}m_{D_{s}}^{2}}{%
m_{c}+m_{s}},  \notag \\
\langle 0|J^{B_{s}^{0}}|B_{s}^{0}(p^{\prime })\rangle &=&\frac{%
f_{B_{s}}m_{B_{s}}^{2}}{m_{b}+m_{s}},
\end{eqnarray}%
where $m_{s}=(93.5\pm 0.8)~\mathrm{MeV}$ is the mass of the $s$ quark. In
computations we use the decay constants of the mesons $f_{D_{s}}=(249.9\pm
0.5)~\mathrm{MeV}$ and $f_{B_{s}}=(234\pm 5)~\mathrm{MeV}$. The vertex $%
\langle B_{s}^{0}(p^{\prime })D_{s}^{+}(q)|\widetilde{H}_{\mathrm{bc}%
}(p,\epsilon )\rangle $ has the form%
\begin{equation}
\langle B_{s}^{0}(p^{\prime })D_{s}^{+}(q)|\widetilde{H}_{\mathrm{bc}%
}(p,\epsilon )\rangle =\widetilde{g}_{3}(q^{2})\epsilon _{\mu \nu
}(p)p^{\prime \mu }q^{\nu }.
\end{equation}%
Here,\ $\widetilde{g}_{3}(q^{2})$ is the form factor that describes the
strong interaction of the particles at the vertex $\widetilde{H}_{\mathrm{bc}%
}B_{s}^{0}D_{s}^{+}$. Then $\widehat{\Pi }_{\alpha \beta }^{\mathrm{Phys}%
}(p,p^{\prime })$ is equal to
\begin{eqnarray}
&&\widehat{\Pi }_{\alpha \beta }^{\mathrm{Phys}}(p,p^{\prime })=\frac{%
\widetilde{g}_{3}(q^{2})\widetilde{\Lambda }%
f_{D_{s}}m_{D_{s}}^{2}f_{B_{s}}m_{B_{s}}^{2}}{(m_{c}+m_{s})(m_{b}+m_{s})%
\left( p^{2}-\widetilde{m}^{2}\right) }  \notag \\
&&\times \frac{1}{(q^{2}-m_{D_{s}}^{2})(p^{\prime 2}-m_{B_{s}}^{2})}\left\{
\frac{\widehat{\lambda }^{2}}{3}g_{\alpha \beta }+\right.  \notag \\
&&\left. +\left[ \frac{m_{B_{s}}^{2}}{\widetilde{m}^{2}}+\frac{2\widehat{%
\lambda }^{2}}{3\widetilde{m}^{2}}\right] p_{\alpha }p_{\beta }+\text{other
terms}\right\} ,
\end{eqnarray}%
where $\widehat{\lambda }=\lambda (\widetilde{m},m_{B_{s}},q)$.

The correlator $\widehat{\Pi }_{\alpha \beta }(p,p^{\prime })$ calculated in
terms of the quark-gluon propagators reads%
\begin{eqnarray}
&&\widehat{\Pi }_{\alpha \beta }^{\mathrm{OPE}}(p,p^{\prime })=\int
d^{4}xd^{4}ye^{ip^{\prime }x}e^{iqy}g_{s}\frac{{\lambda }_{ab}^{n}}{2}%
\widetilde{G}_{\rho \alpha }^{n}(0)  \notag \\
&&\times \mathrm{Tr}\left[ \gamma _{5}S_{s}^{ij}(x-y)\gamma
_{5}S_{c}^{jb}(y)\gamma _{5}\sigma _{\beta }^{\rho }S_{b}^{ai}(-x)\right] .
\end{eqnarray}%
The sum rule for $\widetilde{g}_{3}(q^{2})$ is derived by using the terms $%
g_{\alpha \beta }$ in the correlation functions $\widehat{\Pi }_{\alpha
\beta }^{\mathrm{Phys}}(p,p^{\prime })$ and $\widehat{\Pi }_{\alpha \beta }^{%
\mathrm{OPE}}(p,p^{\prime })$ and corresponding amplitudes%
\begin{eqnarray}
&&\widetilde{g}_{3}(q^{2})=\frac{%
3(m_{c}+m_{s})(m_{b}+m_{s})(q^{2}-m_{D_{s}}^{2})}{\widetilde{\Lambda }%
f_{D_{s}}m_{D_{s}}^{2}f_{B_{s}}m_{B_{s}}^{2}\widehat{\lambda }^{2}}  \notag
\\
&&\times e^{m^{2}/M_{1}^{2}}e^{m_{B_{s}}^{2}/M_{2}^{2}}\widehat{\Pi }(%
\mathbf{M}^{2},\mathbf{s}_{0},q^{2}).  \label{eq:SR3}
\end{eqnarray}

Numerical computations are carried out using Eq.\ (\ref{eq:SR3}) and working
windows Eq.\ (\ref{eq:Wind1A}) for the parameters $(M_{1}^{2},\ s_{0})$. We
employ also regions
\begin{equation}
M_{2}^{2}\in \lbrack 5.5,6.5]~\mathrm{GeV}^{2},\ s_{0}^{\prime }\in \lbrack
34,35]~\mathrm{GeV}^{2}
\end{equation}%
in the $B_{s}^{0}$ channel.

The strong coupling $\widetilde{g}_{3}$ is determined at the mass shell $%
q^{2}=m_{D_{s}}^{2}$of the $D_{s}^{+}$ meson
\begin{equation}
\widetilde{g}_{3}\equiv \widetilde{\mathcal{G}}_{3}(-m_{D_{s}}^{2})=(3.87\pm
0.66)\ \mathrm{GeV}^{-1}.
\end{equation}%
The function $\widetilde{\mathcal{G}}_{3}(Q^{2})$ is fixed by the
coefficients $\widetilde{\mathcal{G}}_{3}^{0}=5.02\ \mathrm{GeV}^{-1},$ $%
\widetilde{c}_{3}^{1}=3.83,$and $\widetilde{c}_{2}^{2}=-2.10$. The relevant
SR predictions for the form factor $\widetilde{g}_{3}(Q^{2})$ and $%
\widetilde{\mathcal{G}}_{3}(Q^{2})$ are plotted in Fig.\ \ref{fig:Fit1}. The
width of the decay $\widetilde{H}_{\mathrm{bc}}\rightarrow
B_{s}^{0}D_{s}^{+} $ is

\begin{equation}
\Gamma \left[ \widetilde{H}_{\mathrm{bc}}\rightarrow B_{s}^{0}D_{s}^{+}%
\right] =(28.8\pm 7.0\pm 4.6)~\mathrm{MeV}.
\end{equation}

The decay $\widetilde{H}_{\mathrm{bc}}\rightarrow B_{s}^{\ast 0}D_{s}^{\ast
+}$ can be explored by the same way. Therefore, we omit details and provide
only numerical values of quantities used in calculations. Thus, the masses
of the mesons $B_{s}^{\ast 0}$ and $D_{s}^{\ast +}$ are $m_{B_{s}^{\ast
}}=(5415.4\pm 1.4)~\mathrm{MeV}$ and $m_{D_{s}^{\ast }}=(2112.2\pm 0.04)~%
\mathrm{MeV}$, respectively. For the decay constants of these particle we
employ $f_{B_{s}^{\ast }}=(221\pm 7)~\mathrm{MeV}$ and $f_{D_{s}^{\ast
}}=(268.8\pm 6.5)~\mathrm{MeV}$. In these computations we make use of the
regions
\begin{equation}
M_{2}^{2}\in \lbrack 6,7]~\mathrm{GeV}^{2},\ s_{0}^{\prime }\in \lbrack
35,36]~\mathrm{GeV}^{2}
\end{equation}%
in the channel of the meson $B_{s}^{\ast 0}$.

The coupling constant $\widetilde{g}_{4}$ at the vertex $\widetilde{H}_{\mathrm{%
bc}}B_{s}^{\ast 0}D_{s}^{\ast +}$ equals to

\begin{equation}
\widetilde{g}_{4}\equiv \widetilde{\mathcal{G}}_{4}(-m_{D_{s}^{\ast
}}^{2})=(0.61\pm 0.12)\ \mathrm{GeV}^{-1},
\end{equation}%
which is found by means of the fit function $\widetilde{\mathcal{G}}%
_{4}(Q^{2})$ with the parameters $\widetilde{\mathcal{G}}_{4}^{0}=0.67\
\mathrm{GeV}^{-1},$ $\widetilde{c}_{4}^{1}=1.18,$and $\widetilde{c}%
_{4}^{2}=-0.96$.

The width of the process $\widetilde{H}_{\mathrm{bc}}\rightarrow B_{s}^{\ast
0}D_{s}^{\ast +}$ amounts to

\begin{equation}
\Gamma \left[ \widetilde{H}_{\mathrm{bc}}\rightarrow B_{s}^{\ast
0}D_{s}^{\ast +}\right] =(25.9\pm 7.2\pm 3.5)~\mathrm{MeV}.
\end{equation}

Decays explored here enable one to evaluate the full width of the exotic
meson $\widetilde{H}_{\mathrm{bc}}$ and find
\begin{equation}
\Gamma \left[ \widetilde{H}_{\mathrm{bc}}\right] =(184.3\pm 22.8)~\mathrm{MeV%
}.
\end{equation}


\section{Summing up}

\label{sec:Dis} 
We have computed the masses\ and widths of the tensor hybrid mesons $H_{%
\mathrm{bc}}$ and $\widetilde{H}_{\mathrm{bc}}$ with spin-parities $J^{%
\mathrm{PC}}=2^{-}$ and $J^{\mathrm{PC}}=2^{+}$, respectively. Results
obtained for these parameters confirm that hybrids $H_{\mathrm{bc}}$ and $%
\widetilde{H}_{\mathrm{bc}}$ can strongly decay to ordinary meson pairs.

We have evaluated the full width of the mesons $H_{\mathrm{bc}}$ and $%
\widetilde{H}_{\mathrm{bc}}$ as well. As decay modes of $H_{\mathrm{bc}}$ we
have explored the processes $H_{\mathrm{bc}}\rightarrow B^{\ast 0}D^{+}$ and
$B^{\ast +}D^{0}$. The full width of $H_{\mathrm{bc}}$ evaluated by taking
into account these two decays amounts to $(50.8\pm 9.8)~\mathrm{MeV}$. The
width of the exotic state $\widetilde{H}_{\mathrm{bc}}$ equals to $(184.3\pm
22.8)~\mathrm{MeV}$ which have been estimated by considering the processes $%
\widetilde{H}_{\mathrm{bc}}\rightarrow B^{+}D^{0}$, $B^{0}D^{+}$, $B^{\ast
+}D^{\ast 0}$, $B^{\ast 0}D^{\ast +}$, $B_{s}^{0}D_{s}^{+}$, and $%
B_{s}^{\ast 0}D_{s}^{\ast +}$. Clearly, the hybrid meson $\widetilde{H}_{%
\mathrm{bc}}$ can be interpreted as a broad structure.

Performed studies have extended our knowledge about yet hypothetical tensor
mesons $H_{\mathrm{bc}}$ and $\widetilde{H}_{\mathrm{bc}}$ and gave us
interesting details of their features. These structures and their decays
deserve investigations also in the context of other methods. This will
allow us to collect valuable information about the heavy hybrid $\overline{b}%
gc$ mesons.  Results of the present article form a sizeable part of such
useful information.

\section*{ACKNOWLEDGMENTS}

K. Azizi gratefully acknowledges the Iran National Science Foundation (INSF) for partial financial support under Elites Grant No. 4037888.

\end{document}